\documentclass[fleqn,twoside]{article}
\usepackage{espcrc2}

\usepackage{graphicx}
\usepackage[figuresright]{rotating}

\newcommand{\AmS}{{\protect\the\textfont2
  A\kern-.1667em\lower.5ex\hbox{M}\kern-.125emS}}

\title{
{
\vspace{-3.95cm} \normalsize \hfill
\parbox{35mm}{DESY 03-139\\MS-TP-03-9\\HU-EP-03164\\SFB/CPP-03-35
\\CERN-TH/2003-211\\
}}\\[15mm]
Static quarks with improved statistical precision
\thanks{Talk~given~by~M.~Della~Morte.~Work supported in part by the 
European Community's Human Potential Programme, contract 
HPRN-CT-2000-00145, Hadrons/Lattice QCD and by the DFG in the SFB/TR 09.}}
\author{M.~Della Morte\address[Zeu]{DESY Zeuthen, Platanenallee 6, 15738
 Zeuthen, Germany}, S.~D\"urr\addressmark[Zeu],
J.~Heitger\address[Mu]{WWU M\"unster, Institut f\"ur Theoretische Physik,
Wilhelm--Klemm-Str. 9, 48149 M\"unster, Germany},
H.~Molke\addressmark[Zeu],
J.~Rolf\address[HU]{Institut f\"ur Physik, Humboldt--Universit\"at zu 
Berlin, Invalidenstr. 110, 10115 Berlin, Germany}, 
A.~Shindler\address[NIC]{NIC/DESY Zeuthen, Platanenallee 6, 15738
  Zeuthen, Germany} and
R.~Sommer\addressmark[Zeu]$^,$\address[Cern]{CERN--TH, CH--1211 Geneva 23, 
Switzerland}\newline
(ALPHA Collaboration)}

\begin{document}

\begin{abstract}
We present a numerical study for different discretisations of the static 
action, concerning  cut-off effects and the growth of statistical errors
with Euclidean time.
An error reduction by an order of magnitude can be obtained with respect
to the Eichten-Hill action, for time separations beyond 1.3 fm,
keeping discretization errors small.
The best actions lead to a big improvement on the precision of the
quark mass $M_{\rm b}$ and $F_{\rm B_s}$ in the static approximation.
\vspace{1pc}
\end{abstract}

\maketitle

\section{INTRODUCTION}
\vspace{-0.1cm}
Euclidean correlation functions in the static approximation are known to
be very noisy, the noise to signal ratio $R_{\rm NS}$ growing exponentially 
with the time separation.  For the Eichten-Hill (EH) action~\cite{eh}, the law 
\begin{equation}
 R_{\rm NS}\equiv {\rm noise \over \rm signal} 
  \propto \exp\left(x_0\,\Delta\right)\,, \; \Delta=E_{\rm stat}-m_{\pi},
\label{rns}
\end{equation}
is roughly fulfilled~\cite {hashi},  with the ground state energy of a heavy 
meson $E_{\rm stat}$ being linearly divergent while approaching the continuum
 limit.

Here we explore the possibility of reducing the exponent in eq.~(\ref{rns})
by changing the discretisation of the static lattice action. On the 
other hand we want to  retain some properties of the Eichten-Hill
action in order to 
preserve the same level of O$(a)$ improvement, in particular
\begin{itemize}
\item heavy quark spin symmetry,
\item local conservation of heavy quark flavor number,
\end{itemize}
together with gauge, cubic and parity invariance and locality. Writing the 
static lattice action
\begin{eqnarray}
  S_{\rm stat}^{\rm W} \!\!\!\!\! &=& \!\!\! a^4  
            \sum_x \overline{\psi}_{\rm h}(x) D_0 \psi_{\rm h}(x), \\
     D_0 \psi_{\rm h}(x) \!\!\!\!\! &=& \!\!\!
 {1\over a} [\psi_{\rm h}(x) - W^\dagger(x-a\hat{0},0)\psi_{\rm h}
  (x-a\hat{0})], \nonumber
\end{eqnarray}
we studied the cases
\begin{eqnarray}
&& W_{\rm S}(x,0) = 
 V \left[{g_0^{2} \over 5} \!\!+\!\! 
                \left({1\over 3} {\rm tr} V^\dagger V\!\!
                \right)^{1/2}\right]^{-1}\!\!\!\,,  \\
&&  W_{\rm A}(x,0) = V\,, \\
&& W_{\rm HYP}(x,0) = V_{\rm HYP}\,, 
\end{eqnarray}
where $V$ is the average of the 6 staples around the link $U(x,0)$ and 
$V_{\rm HYP}$ is the HYP-link~\cite{hyp}.\footnote{For the coefficients
$\alpha_1,\alpha_2$ and $\alpha_3$ defining the HYP-smearing we used the values
proposed in~\cite{hyp}.}
\vspace{-0.2cm}
\section{NUMERICAL RESULTS}
\vspace{-0.1cm}
We mainly looked at the gain concerning noise reduction and at some scaling 
properties for the new actions. 
The setup is defined by the Schr\"odinger functional scheme implemented with
 non-perturbatively O($a$) improved Wilson actions for the gauge and
the light quark sectors.
A more detailed discussion of the framework and 
definitions of the correlation functions can be found in~\cite{impHQET}.
Our aim is the computation of the $B_{\rm s}$-meson decay constant as
well as other b-physics matrix elements.
\subsection{Noise reduction and cut-off effects}
\begin{figure}[htb]
\vspace{-0.55cm}
\includegraphics[width=18pc]{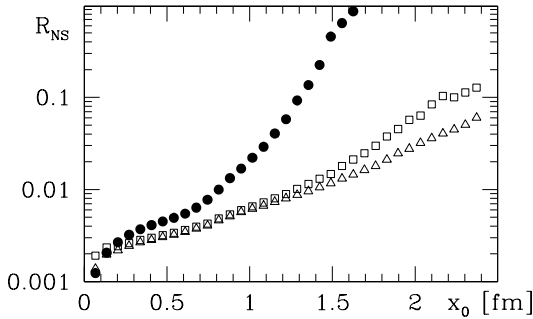}
\vspace{-1.37cm}
\caption{$R_{\rm NS}$ for $f_{\rm A}^{\rm stat}$ 
from a $24^3 \times 36$, $\beta=6.2$ lattice.
See text for the symbols.}
\label{rnsf}
\vspace{-0.8cm}
\end{figure}

In Fig.~\ref{rnsf} we show $R_{\rm NS}$ for the static-light axial correlator 
$f_{\rm A}^{\rm stat}$ as obtained from an ensemble of 2500 quenched
configurations.
This is one of the ensembles used for the
computation of $F_{\rm B_{\rm s}}\sqrt{m_{\rm B_{\rm s}}}$ in the static limit.
The Figure shows that more than an order of magnitude can be gained in 
$R_{\rm NS}(x_0 \simeq 1.5$ fm) 
with respect to the EH action (filled circles) by using  
$S_{\rm stat}^{\rm S}$  (empty squares), which behaves similarly to
$S_{\rm stat}^{\rm A}$. The picture is even better for 
$S_{\rm stat}^{\rm HYP}$ (empty triangles).

\begin{figure}[htb]
\vspace{-0.05cm}
\includegraphics[width=18pc]{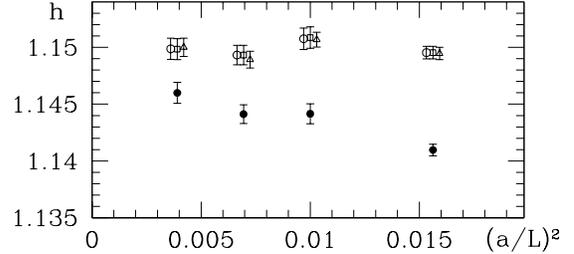}
\vspace{-1.37cm}
\caption{Scaling plot for $h$. Symbols as in Fig.~\ref{rns} (empty
  circles refer to $S_{\rm stat}^{\rm A}$).
}
\label{rat}
\vspace{-0.8cm}
\end{figure}

We then studied the scaling behaviour for a set of observables. 
In~\cite{impHQET} we reported
about the step scaling function of the static-light axial
current renormalisation constant $Z_{\rm A}^{\rm stat}$
(see~\cite{zastat}).
Here we want to present a study of the quantity
\begin{equation}
h={{\left.f_{\rm A}^{\rm stat}(T/2)\right |_{\theta=0}}
	\over
	 {\left.f_{\rm A}^{\rm stat}(T/2)\right |_{\theta=1/2}}}\quad
	\mbox{at}\quad  {m_{\rm q}=0},
\label{h}
\end{equation}
where $\theta$ is the angle defining the periodicity of the
fermions.
Our results are shown in Fig.~\ref{rat}, they refer to four different lattice
resolutions of an $L^4$ volume with $L=1.436r_0$ with $r_0=0.5$ fm.

\begin{figure}[htb]
\vspace{-0.05cm}
\includegraphics[width=18pc]{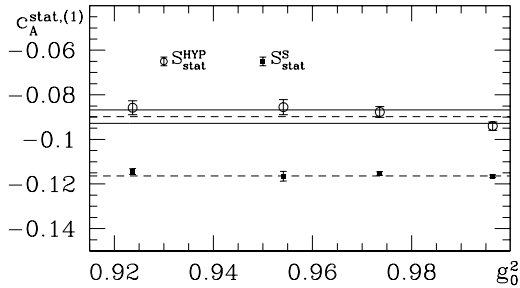}
\vspace{-1.4cm}
\caption{Numerical results for $c_{\rm A}^{\rm stat(1)}$. The dashed
  line in the lower part is the analytic perturbative result and the error
  band in the upper part is our estimate for $c_{\rm A}^{\rm stat}$ for 
$S_{\rm stat}=S_{\rm stat}^{\rm HYP}$.
}
\label{ca}
\vspace{-0.7cm}
\end{figure}

Throughout all our computations  we defined the O($a$)
improved correlator $f_{\rm A}^{\rm stat}$ by using the tree
level value $1/2$ for $b_{\rm A}^{\rm stat}$~\cite{castat} 
and the 1-loop results for $c_{\rm A}^{\rm stat}$.
In particular we set
\begin{eqnarray}
  c_{\rm A}^{\rm stat} \!\!\!\!\! &=& \!\!\!\! -0.08237 g_0^2 + {\rm
  O}(g_0^4), \quad S_{\rm stat}=S_{\rm stat}^{\rm EH}, \nonumber \\
  c_{\rm A}^{\rm stat}  \!\!\!\!\! &=& \!\!\!\! -0.1164(10) g_0^2 \! + \!
  {\rm O}(g_0^4),\, S_{\rm stat}=\! S_{\rm stat}^{\rm S},S_{\rm stat}^{\rm A},
  \nonumber \\
  c_{\rm A}^{\rm stat} \!\!\!\!\! &=& \!\!\!\! -0.090(3) g_0^2 + {\rm
  O}(g_0^4),  \!\quad S_{\rm stat}=S_{\rm stat}^{\rm HYP}. \nonumber
\end{eqnarray}
The first two numbers have been worked out in perturbation theory,
while the last one has been numerically estimated by solving with respect to 
$c_{\rm A}^{\rm stat}(S_{\rm stat}^{\rm HYP})$ the implicit equation
\begin{equation}
{\left.{f_{\rm A}^{\rm stat}(T/2)}\right|^{\rm A}_\theta\over
 \left.{f_{\rm A}^{\rm stat}(T/2)}\right|^{\rm A}_{\theta'}} \,
{\left.{f_{\rm A}^{\rm stat}(T/2)}\right|^{\rm HYP}_{\theta'}\over
 \left.{f_{\rm A}^{\rm stat}(T/2)}\right|^{\rm HYP}_\theta}=1.
\end{equation}
This defines an improvement condition for a discretisation
($S_{\rm stat}^{\rm HYP}$),
once the improvement coefficients for another discretisation
($S_{\rm  stat}^{\rm A}$) are known. In Fig.~\ref{ca} we show our numerical 
results
for the coefficients 
$c_{\rm A}^{\rm stat,(1)}(S_{\rm  stat}^{\rm S})$ 
and $c_{\rm A}^{\rm stat,(1)} (S_{\rm stat}^{\rm HYP})$, taking as input 
$c_{\rm A}^{\rm stat}(S_{\rm stat}^{\rm A})$.
The same data sets used for computing $h$ in eq.~(\ref{h}) have been
used here ($\theta=0$, $\theta'=1/2$). The lower part of the plot can
be regarded as a test of the method.
\subsection{The $B_{\rm s}$-meson decay constant}
The setup has been used for computing 
\begin{equation}
\Phi_{\rm RGI} \propto Z_{\rm RGI} 
  { {f_{\rm A}^{\rm stat}(x_0)}\over{\sqrt{f_1}} } e^{(x_0-T/2) 
E_{\rm stat}(x_0)},
\label{phirgi}
\end{equation}
for which we have calculated the regularisation dependent part of the
renormalisation constant $Z_{\rm RGI}$ exactly as done in~\cite{zastat}
for the EH action. 
Here the $B_{\rm s}$-meson boundary to boundary
correlation function $f_1$ enters. 
The proportionality symbol in eq.~(\ref{phirgi}) summarises
volume factors coming from the normalisation of the correlation
functions, cmp.~\cite{impHQET}. In addition we
introduced hydrogen-like wavefunctions on the boundaries of the
Schr\"odinger functional in order to minimise the overlap with the
first excited state; we arrive at a plateau of length $0.8$~fm in
$\Phi_{\rm RGI}$ and the desired ground state matrix element
can be extracted with confidence.

The quantity $F_{\rm B_{\rm s}}\sqrt{m_{\rm B_{\rm s}}}$ is
related to $\Phi_{\rm RGI}$ via
\begin{equation}
\Phi_{\rm RGI}= { {F_{\rm B_s}\sqrt{m_{\rm B_s}}} \over 
{C_{\rm PS}(M_{\rm b}/\Lambda_{\overline{MS}})}} + O(1/M_{\rm b}),
\end{equation}
the function $C_{\rm PS}(M_{\rm b}/\Lambda_{\overline{MS}})$ \cite{zastat}
being
known in perturbation theory up to and including $\bar{g}^4(m_{\rm b})$
corrections to the leading order~\cite{cit}.

\begin{figure}[htb]
\vspace{-0.05cm}
\includegraphics[width=18pc]{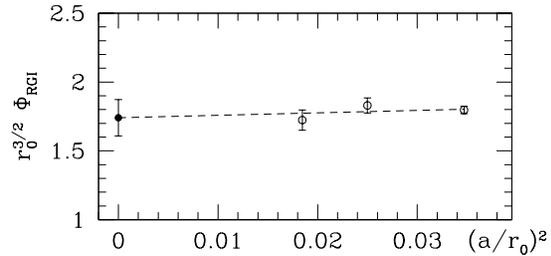}
\vspace{-1.42cm}
\caption{Continuum limit 
  of $\Phi_{\rm RGI}r_0^{3/2}$.
}
\label{phi}
\vspace{-0.85cm}
\end{figure}

We computed $\Phi_{\rm RGI}$ for three
different lattice resolutions ($\beta=6, 6.1$ and $6.2$) on a $L^3 \times
T$ topology with $T/L=3/2$ and $L=$ 1.5 to 1.9 fm. 
The continuum limit extrapolation
of $r_0^{3/2}\Phi_{\rm RGI}$, evaluated from $S_{\rm stat}^{\rm HYP}$ 
at $x_0=T/2$,
is shown in Fig.~\ref{phi}. The
extrapolated value is 
\begin{equation}
r_0^{3/2}\Phi_{\rm RGI}=1.74(13).
\label{Fb}
\end{equation}
\vspace{-0.4cm}
\section{CONCLUSIONS}
\vspace{-0.1cm}
We have shown that the problem with the statistical precision of
correlation functions computed with the Eichten-Hill action can be
overcome by changing the lattice discretisation. In particular the
largest gain is obtained by making use of the HYP-links in the static action.
Cut-off effects for the proposed actions turned out to be of the same
size as for the EH action, and are in general rather small.
The improvements presented led to a computation of the quantity
$\Phi_{\rm RGI}$ in the continuum limit with an error of 7\%. 
We are presently reducing this error further 
by adding one more lattice resolution 
($\beta=6.45$).
The same data can be used for improving the results on the renormalisation
group invariant b-quark mass along the lines described
in~\cite{Mb}. Finally, a precise determination of $F_{\rm B_{\rm s}}$
can be obtained by combining the result in eq.~(\ref{Fb}) with data
around the charm quark mass~\cite{And,jur}.

{\bf Acknowledgements}. We thank NIC/DESY Zeuthen for allocating computer time 
on the APEmille machines to this project, and the APE group for their support.

\end{document}